\documentclass[preprint,12pt]{elsarticle}

\usepackage{amssymb}
\usepackage{amsmath}
\usepackage{graphicx}
\usepackage{subcaption}
\usepackage{placeins} 
\usepackage{array}
\usepackage{amsmath}
\usepackage{multirow} 
\usepackage{booktabs}   
\usepackage{amsmath}   
\usepackage{fancyhdr}
\usepackage{float} 
\usepackage{url}

\journal{NIMA}

\begin{document}

\begin{frontmatter}



\title{4H-SiC PIN detector for alpha particles from  room temperature to 90 °C}


\author[label1,label2]{Xingchen Li} 

\author[label1]{Sen Zhao} 

\author[label1]{Mengke Cai} 

\author[label3]{Suyu Xiao} 

\author[label1]{Congcong Wang} 

\author[label2]{Weimin Song} 

\author[label1]{Xin Shi} 

\author[label1]{Xiyuan Zhang} 

\affiliation[label1]{organization={Institute of High Energy Physics, Chinese Academy of Sciences},
            addressline={19B Yuquan Road, Shijingshan District}, 
            postcode={100049}, 
            state={Beijing},
            country={China}}
\affiliation[label2]{organization={Jilin University},
            addressline={No.2699, Qianjin Road}, 
            city={Changchun},
            postcode={130015}, 
            state={Jilin},
            country={China}}

\affiliation[label3]{organization={Shandong Institute of Advanced Technology},
            addressline={NO.1501, Panlong Road}, 
            city={Jinan},
            postcode={250100}, 
            state={Shandong},
            country={China}}

\begin{abstract}

In the field of high-energy particle detection, detectors operating in high-radiation environments primarily face high costs associated with power consumption and cooling systems. Therefore, the development of particle detectors capable of stable operation at room temperature or even elevated temperatures is of great significance. Silicon carbide (SiC) exhibits significant potential for particle detector applications due to its exceptional carrier mobility, radiation hardness, and thermal stability. Over the past decade, significant breakthroughs in silicon carbide epitaxial growth technology and device processing techniques have enabled the development of SiC-based particle detectors, providing a new technological pathway for particle detection in high-temperature environments.

In this work, we fabricate a 4H-SiC PIN detector, named SIlicon CARbide (SICAR) and characterize its leakage current, capacitance, and charge collection across varying temperatures. The results indicate that the detector maintains a very low leakage current (< 10 nA) at 90 °C, with no degradation in depletion capacitance or charge collection performance. Additionally, it achieves a fast rise time of 333 ps at 90 °C, confirming its potential for high-temperature radiation detection applications.
\end{abstract}




\end{frontmatter}



\section{Introduction}
Silicon-based semiconductor particle detectors have advanced significantly over the past four decades and have been widely used in collider experiments\cite{ITk_TDR}, neutrino detection\cite{juno}, and other high energy physics applications. However, conventional silicon detectors are typically limited to room-temperature or cryogenic operation, rendering them unsuitable for extreme environments such as nuclear reactor monitoring\cite{Nuclear_Heating_Measurements}, where thermal stability is critical.

Silicon carbide (SiC), as a wide-bandgap semiconductor material, exhibits not only high electron mobility\cite{mobility} and strong radiation resistance\cite{radiation_hard1}, but also a thermal conductivity over three times higher than that of silicon. Furthermore, silicon carbide maintains extremely low leakage current levels even at elevated temperatures, making it particularly suitable for harsh-environment applications \cite{High_temperature_electronics} \cite{Wide_Bandgap_Semiconductors_review}. 

Research has demonstrated that SiC Schottky diodes exhibit excellent high-temperature operational stability\cite{high_T_SiC_Neutrons} \cite{high_T_SiC_MIP}. In comparison, SiC PIN devices feature a wider depletion region, higher breakdown voltage, and lower leakage current characteristics. Motivated by these advantages, we fabricated SiC PIN devices and systematically investigated their high-temperature performance.

\section{Device Configuration and Processing Technology}
The 4H-SiC PIN devices under investigation, named SICAR (SIlicon CARbide), were independently developed by the Institute of High Energy Physics (IHEP) at the Chinese Academy of Sciences (CAS).\cite{taoyang_sicar}. The PIN detector employs a fully epitaxial-grown structure, comprising a lightly doped N- active region and a heavily doped P+ contact layer. The P++ contact layer has a thickness of $\rm 0.6 ~ \mu m$ with a doping concentration $\rm > 1 \times 10^{19} ~ cm^{-3}$. Heavy doping facilitates the formation of ohmic contacts, effectively eliminating the Schottky barrier, reducing device conduction losses, and enhancing carrier collection efficiency. The device features an etched termination structure with a 45° $ \sim $  60° etching angle and $1.6 ~ \mu m$ etch depth. A $\rm SiO_2$ passivation layer was deposited via plasma-enhanced chemical vapor deposition (PECVD) at 350 °C. Ohmic contacts were formed by electron beam evaporation of Ni/Ti/Al (50/15/80 nm) multilayer electrodes on both the P++ layer and C-face of the n-type substrate. The size of sensor is $\rm 1.2 ~ mm \times 1.2 ~ mm$, including the field plate structure.

\begin{figure}[H]
  \centering
  \begin{subfigure}[b]{0.45\textwidth}
    \centering
    \includegraphics[width=\textwidth]{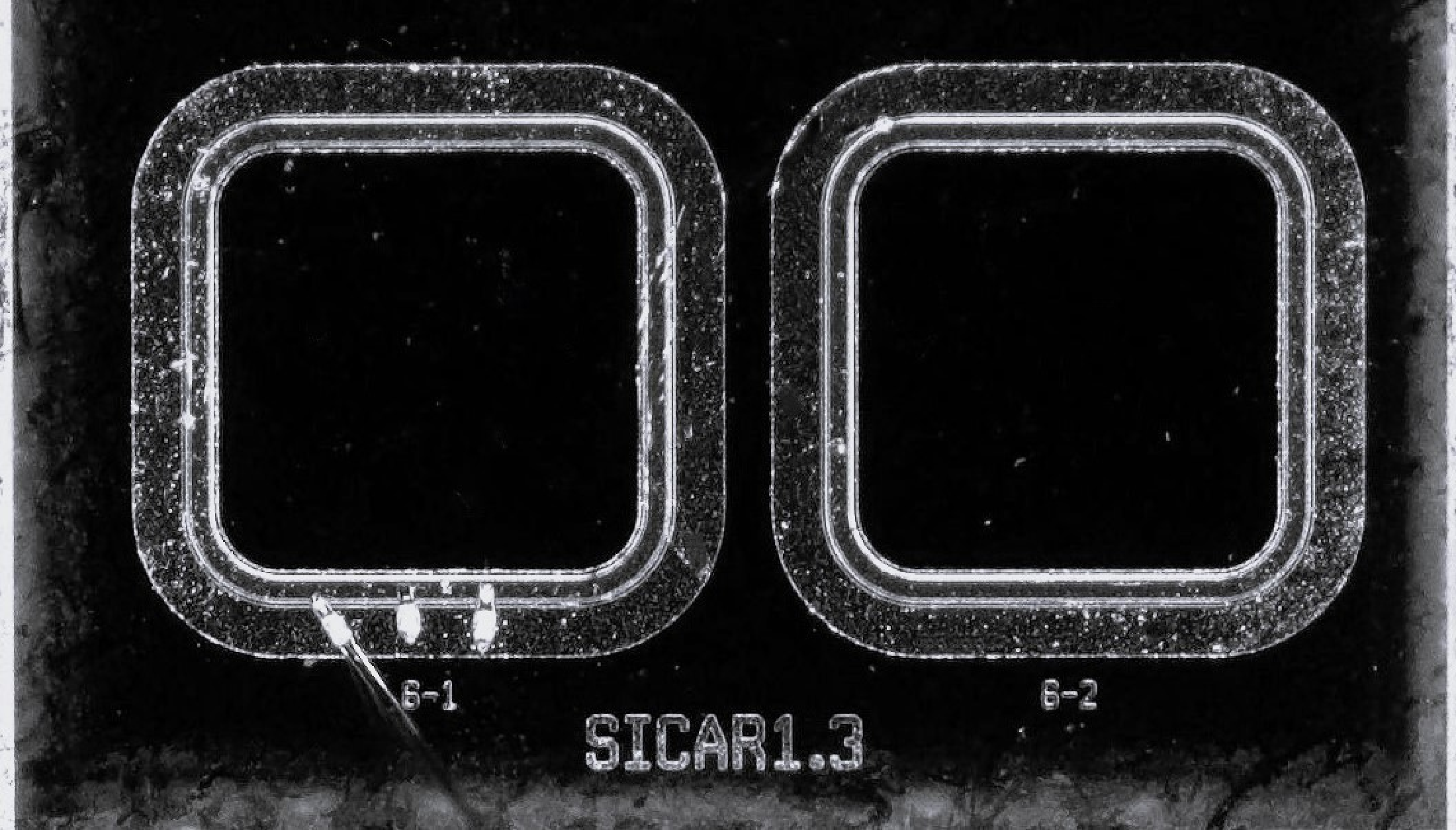} 
    \caption{}
  \end{subfigure}
  \hfill 
  \begin{subfigure}[b]{0.45\textwidth}
    \centering
    \includegraphics[width=\textwidth]{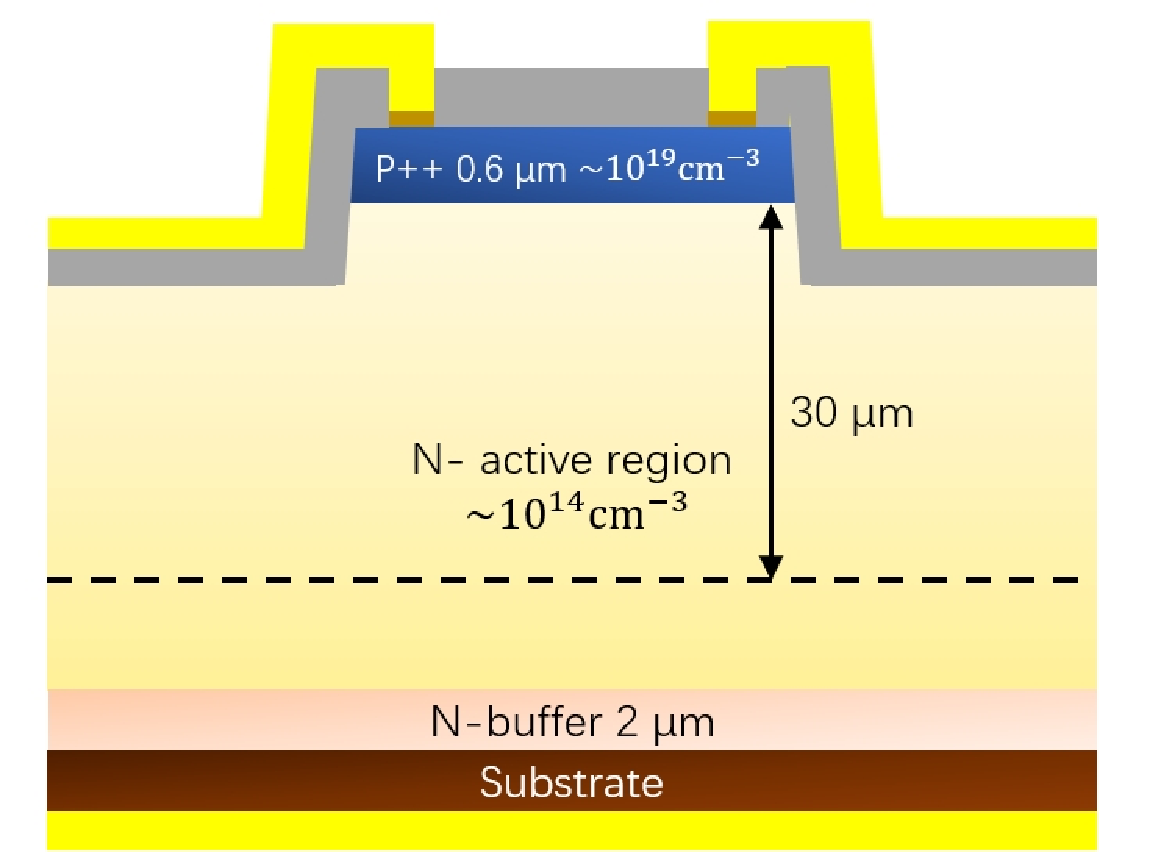} 
    \caption{}
  \end{subfigure}
  \caption{SICAR v1.3 PIN (a) image and (b) epitaxial structure}
  \label{SICAR}
\end{figure}

Specifically, to prevent premature breakdown caused by excessive electric fields at the electrode edges, a field plate structure design using aluminum was incorporated into the electrode structure (Figure \ref{SICAR}(b)). Additionally, to facilitate future laser TCT scanning tests, the electrodes were designed in a ring structure.

The formation of high-quality ohmic contacts is a critical step in the fabrication of PIN devices. For p-type SiC, forming stable ohmic contacts with low resistivity still remains challenging, with ideal ohmic contact resistivity should reach an order of $\rm 1 \times 10^{-5}  ~ \Omega \cdot cm^{2}$\cite{ohmic_contact}. To form high-quality ohmic contacts, appropriate annealing condition is essential. In this work, the Transmission Line Method (TLM)\cite{TLM} is used to evaluate ohmic contact quality through contact resistivity measurements. 

\begin{figure}[H]
\centering
\begin{subfigure}[b]{0.48\textwidth}
\centering
\includegraphics[width=\textwidth]{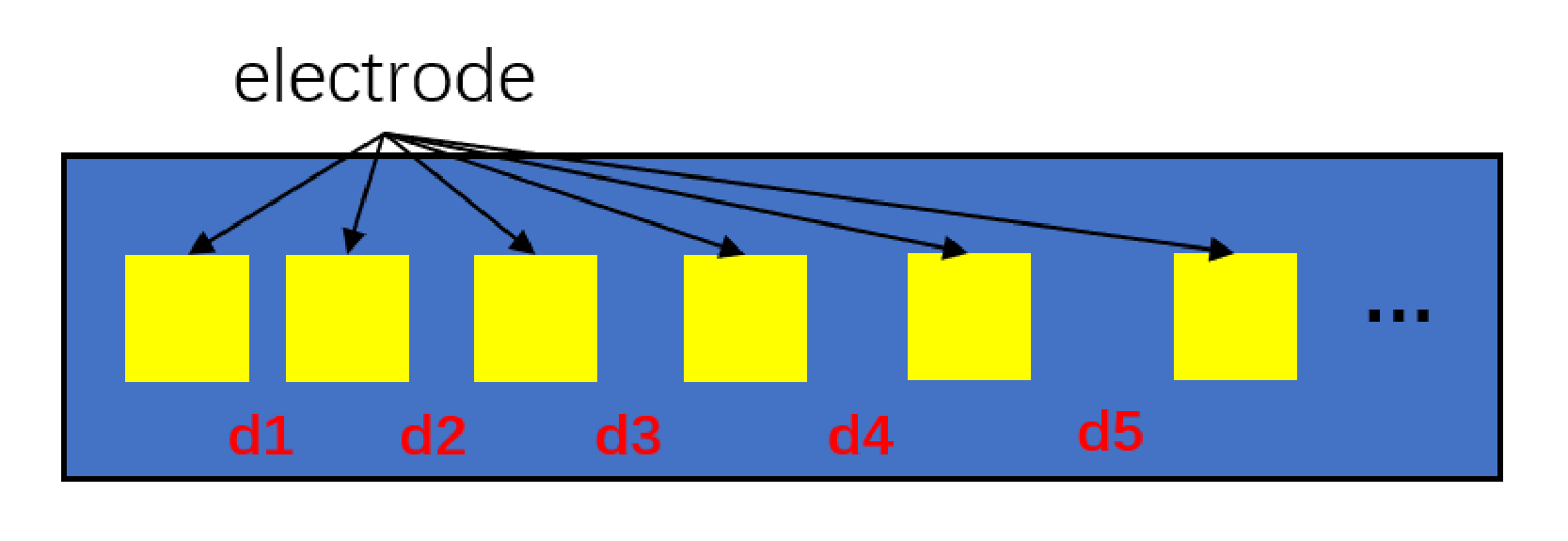} 
\caption{}
\end{subfigure}
\hfill 
\begin{subfigure}[b]{0.48\textwidth}
\centering
\includegraphics[width=\textwidth]{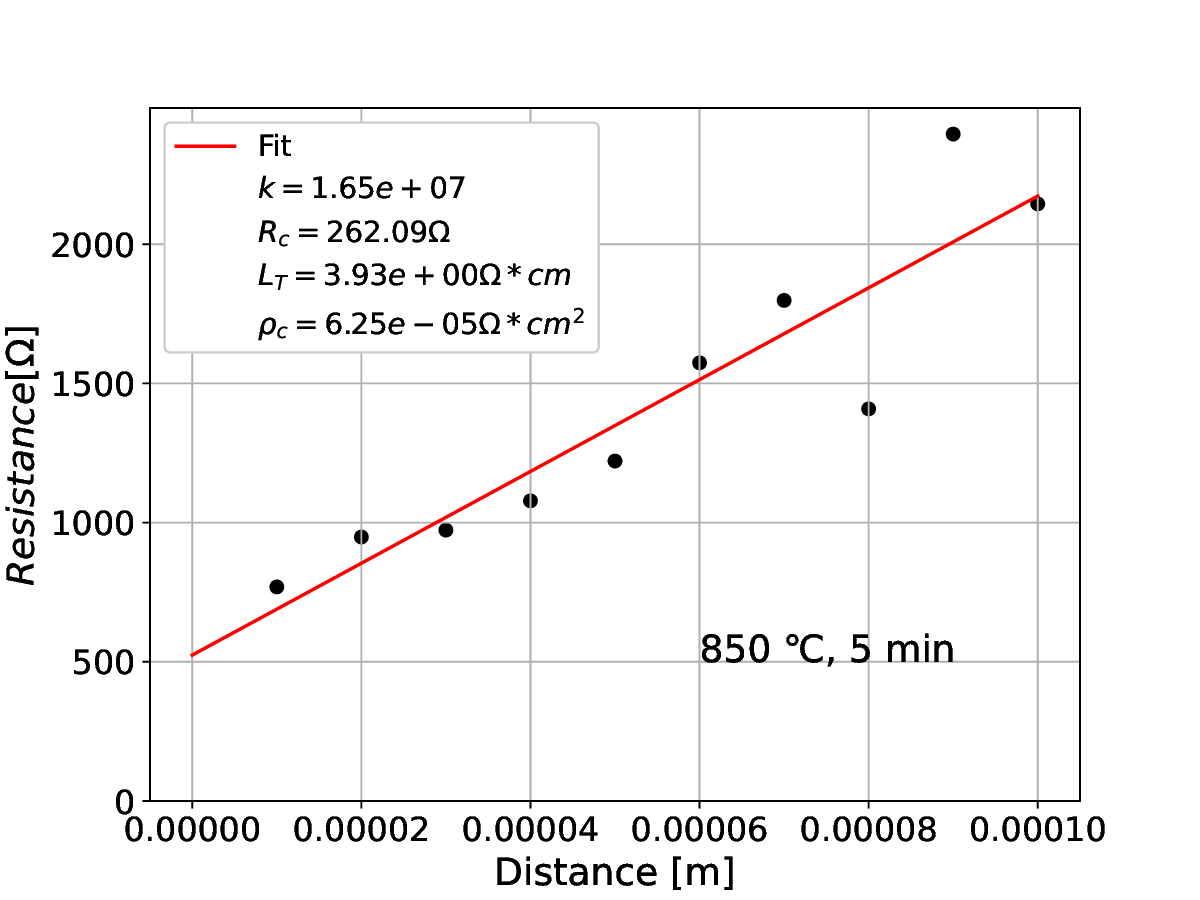} 
\caption{}
\end{subfigure}
\caption{using TLM method get contact resistivity at 850 ℃ 5 min annealing conditions}
\label{contact_resistivity}
\end{figure}

Figure \ref{contact_resistivity} (a) illustrates the contact resistivity testing sample structure. The structure features electrodes with size of $150 ~ \mu m \times 150 ~ \mu m$, spaced at varying distances d ranging from 10 $\mu m$ to 100 $\mu m$. The contact resistivity $\rho_c$ can be calculated using the following formula:

\begin{equation}
\rho_c= \frac{R_c \cdot R_c \cdot A}{R_{sh}}
\end{equation}

where $R_c$ is the contact resistance of the electrode, $A$ is the electrode area and $R_{sh}$ is the sheet resistance of the electrode. The values of $R_c$ and $R_{sh}$ can be determined by fitting the measured resistance values between different distance electrodes:

\begin{equation}
R = 2R_c + R_{sh}\frac{d}{w}
\end{equation}

where d is the distance between the electrodes, and w is the side length of the electrode.

The annealing temperature was scanned from 750 °C to 950 °C with different annealing time. Among them, the 850 °C sample’s IV curves exhibited best linear behavior. As Figure \ref{contact_resistivity} (b) shown, the sample after annealing at 850 °C for 5 minutes have very low contact resistivity which reach $\rm 6.25 \times 10^{-5} ~ \Omega \cdot cm^{2}$, satisfying the requirements for PIN device fabrication.

\section{Experiment setup}
For IV and CV measurement, a specialized test board was utilized instead of conventional probes to achieve better control over environmental temperature and humidity. The measurement system was housed in a controlled chamber maintaining humidity below 5$\%$, with temperature stabilization at five set points: 23 °C, 30 °C, 50 °C, 70 °C, and 90 °C. A Keithley 2470 was used to apply bias voltage for both IV, CV measurement and readout leakage current data. Using Keysight E4980A LCR to add AC voltage and readout capacitance under CV measurement. The measurement protocol employed a 2 V step voltage, with each bias point measured five times to obtain an averaged value, ensuring data reliability.

The charge collection measurement was conducted by using Americium-241\cite{Am241} source. The test environment was maintained humidity levels below 5$\%$, while temperature was varied from 23 °C to 90 °C. A UCSC single-channel board\cite{ucsc_board} was used to readout, which can provide sufficient bandwidth (0 - 2.5 GHz) and signal-to-noise ratio (SNR) for the measurements. Keithley 2470 was used as high voltage source to bias sensor and GWinstek GPD-3303S was used as low voltage source to supply board's power. For oscilloscope, the tektronix MSO64 was used, bandwidth was set to 2.5 GHz, impedance was set to $50 ~ \Omega$ and sampling rate was 10 GS/s.

\section{Measurement results}

\begin{figure}[H]
  \centering
  \begin{subfigure}[b]{0.49\textwidth}
    \centering
    \includegraphics[width=\textwidth]{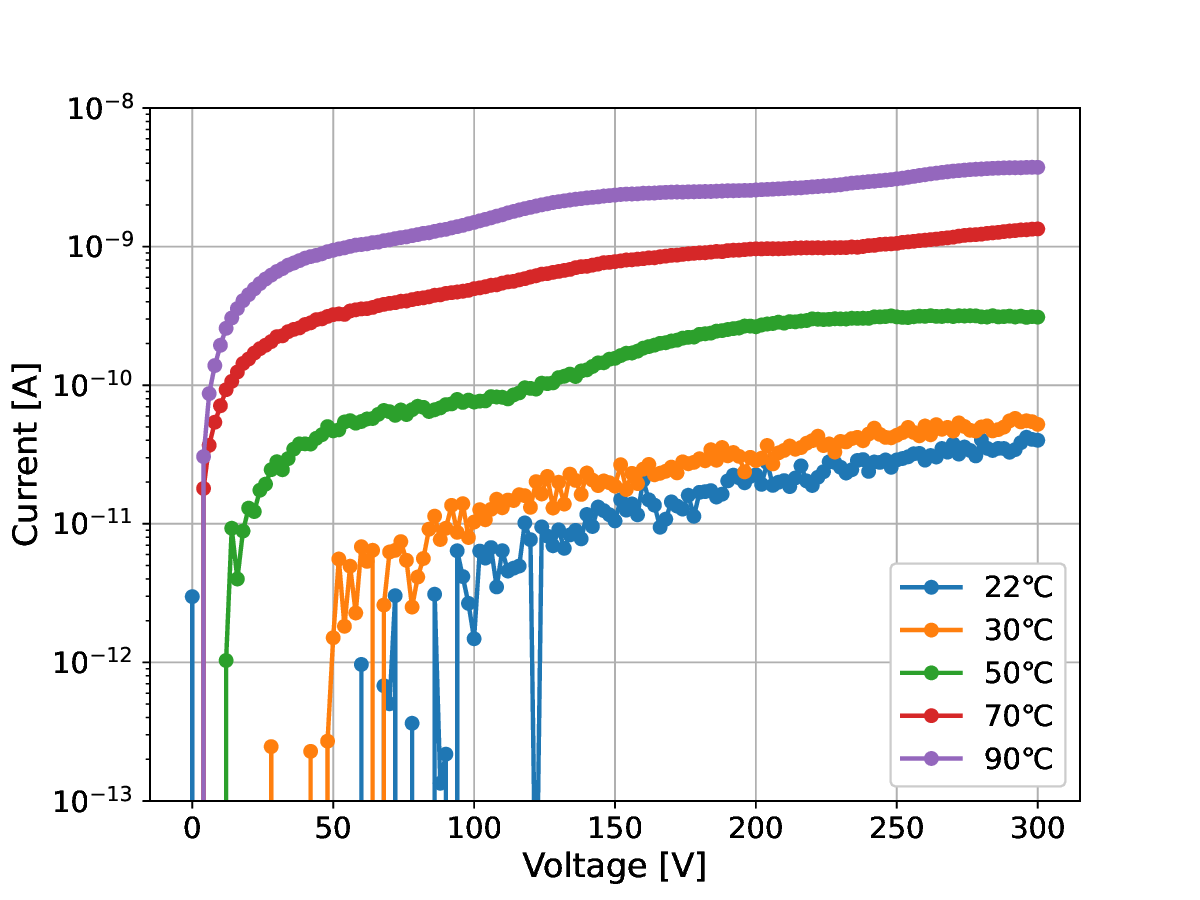} 
    \caption{}
  \end{subfigure}
  \hfill 
  \begin{subfigure}[b]{0.49\textwidth}
    \centering
    \includegraphics[width=\textwidth]{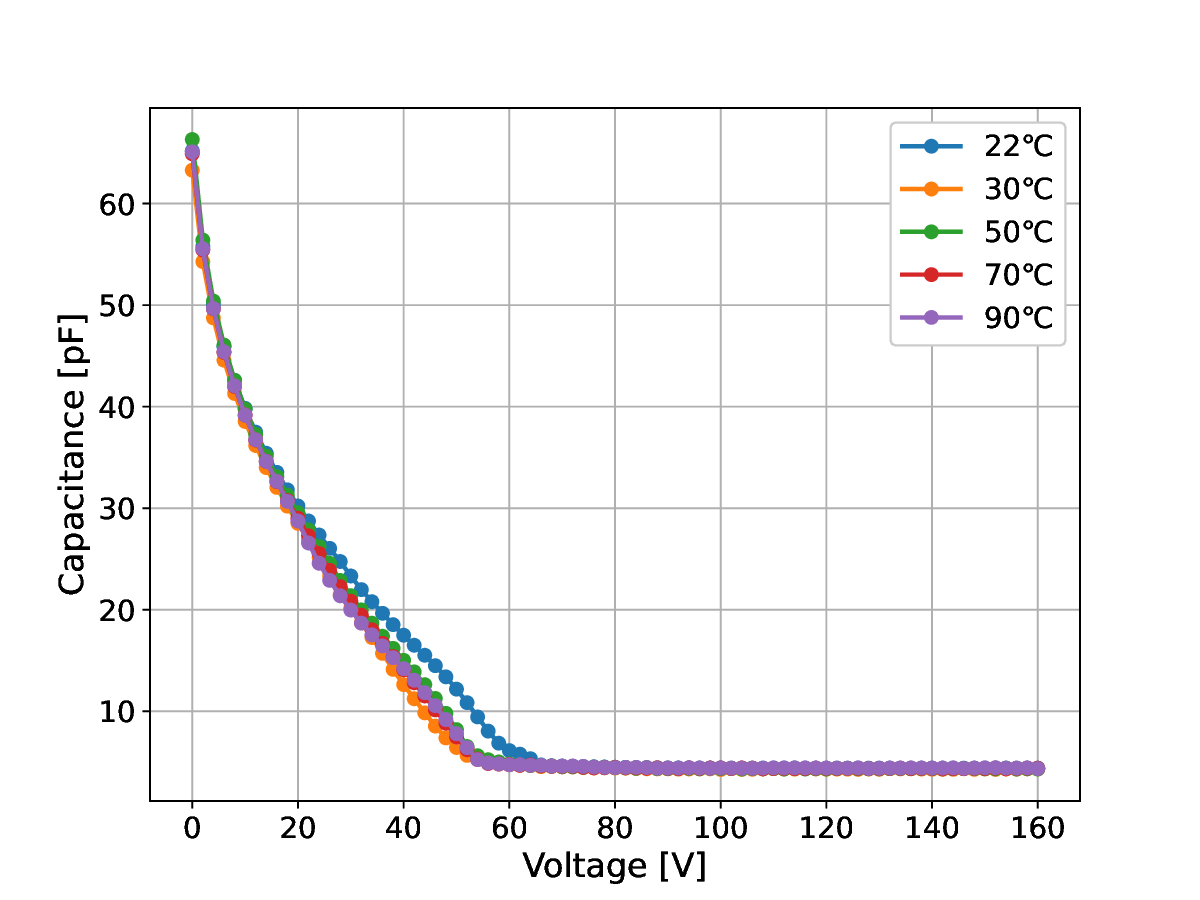} 
    \caption{}
  \end{subfigure}
  \caption{(a) leakage current and (b) capacitance with different temperature}
  \label{IVCV}
\end{figure}

Figure \ref{IVCV} (a) shows the results of the leakage current versus the
reverse voltage bias with different temperature. The experimental results demonstrate that the SICAR PIN sensor exhibits an extremely low leakage current level(< 0.1 nA) at room temperature, and maintains leakage currents below 10 nA at 300 V bias, even at elevated temperatures up to 90 °C. This outstanding performance indicates excellent signal-to-noise ratio (SNR) preservation under high-temperature operating conditions.

Figure \ref{IVCV} (b) shows the results of CV measurement with different temperature under 10 kHz. As illustrated in the figure, temperature exhibits no influence on the depletion capacitance of the device (> 80 V), with only minimal variations in pre-depletion capacitance characteristics. The device achieves an exceptionally low depletion capacitance of 4.5 pF, demonstrating its excellent high-frequency performance capabilities. The CV measurement also reveals a consistent depletion width of approximately 30 $\mu m$ at different temperature.

\begin{figure}[H]
\centering
\includegraphics[width=0.8\textwidth]{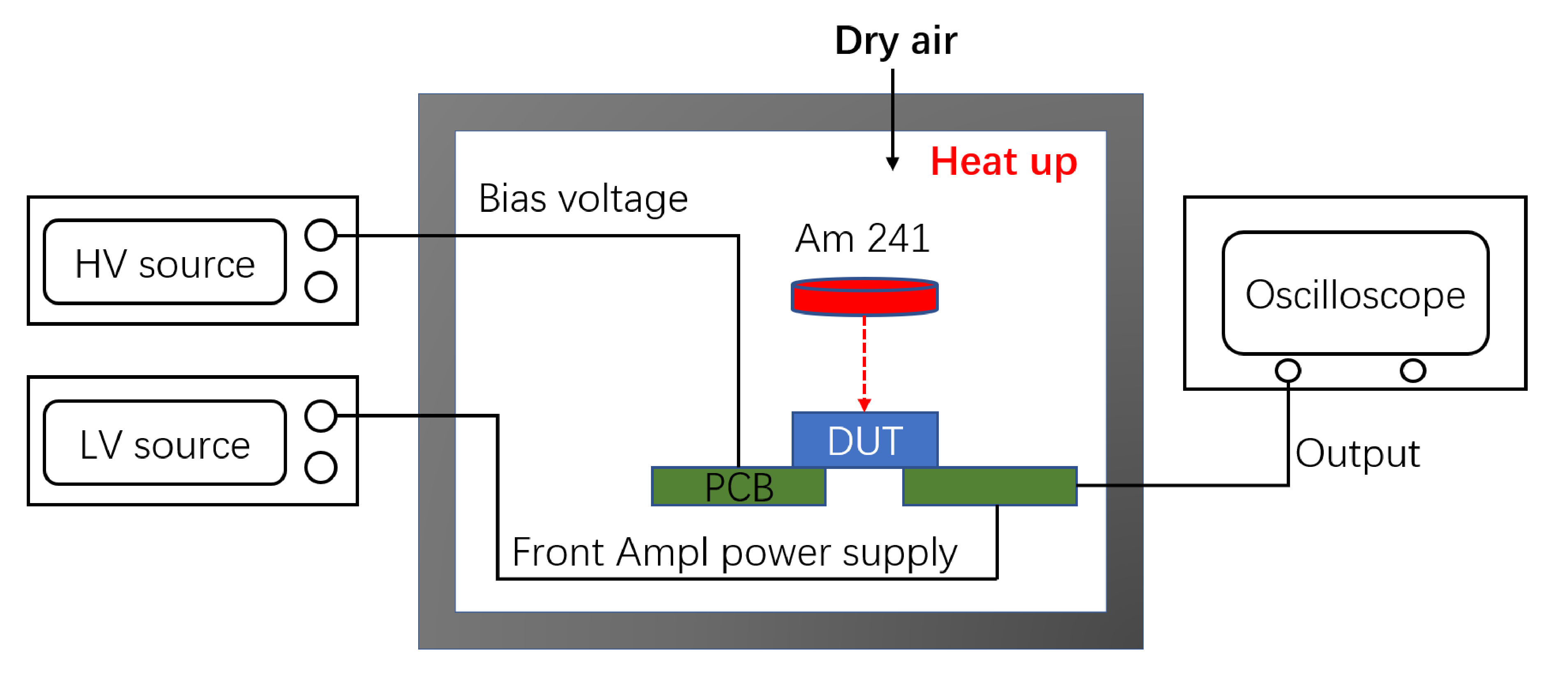}
\caption{$\alpha$ source CCE setup}
\label{alpha_setup}
\end{figure}

Figure \ref{alpha_setup} illustrates a schematic layout for charge collection measurement setup, with the alpha source positioned approximately 7 mm from the device.

\begin{figure}[H]
  \centering
  \begin{subfigure}[b]{0.49\textwidth}
    \centering
    \includegraphics[width=\textwidth]{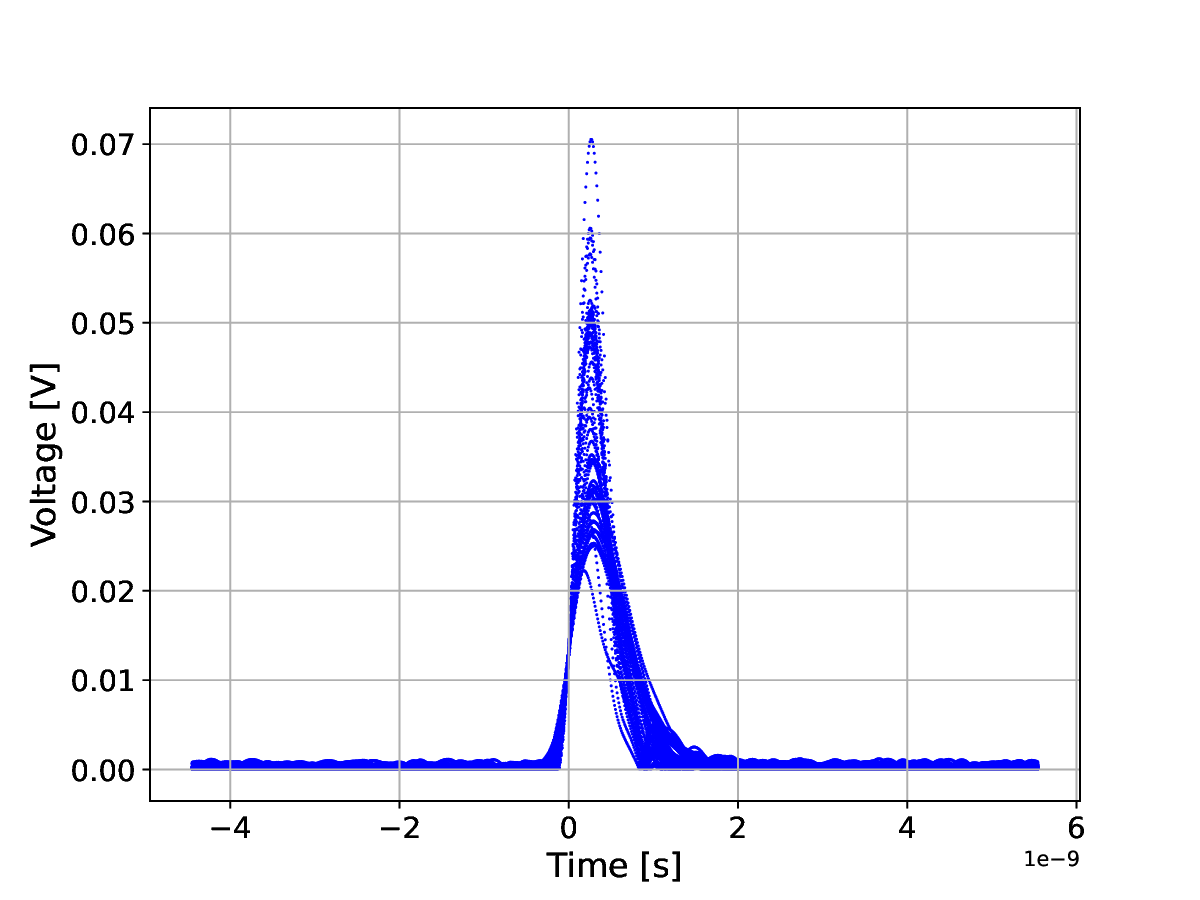} 
    \caption{}
  \end{subfigure}
  \hfill 
  \begin{subfigure}[b]{0.49\textwidth}
    \centering
    \includegraphics[width=\textwidth]{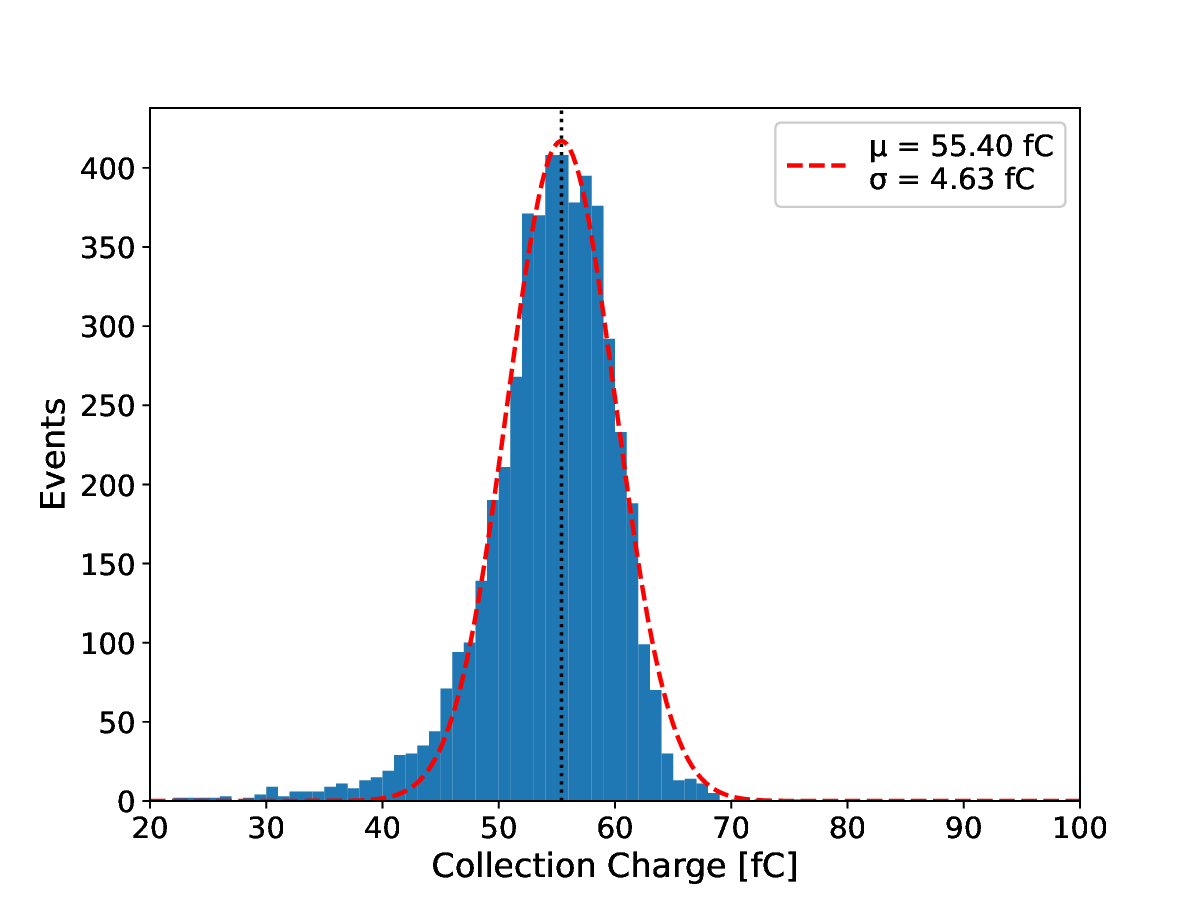} 
    \caption{}
  \end{subfigure}
  \caption{23 ℃, 300 V, $\alpha$ source test signal (a) waveform and (b) charge collection}
  \label{waveform}
\end{figure}

Figure \ref{waveform} (a) the charge collection signal waveform, with the pulse broadening is smaller than 2 ns, demonstrating the advantage of 4H-SiC's high carrier mobility. The significant signal amplitude variations attributed to the non-uniform weighting field distribution of the ring electrode structure.

The measured voltage signals were converted to current signals using the board's calibrated transimpedance gain. Figure \ref{waveform} (a) displays the charge collection distribution obtained through signal integration at 23 °C under 300 V bias, with Gaussian fitting applied to extract the mean value and variance. While the majority of events follow a Gaussian distribution, a distinct low-charge tail is observed, attributable to the Landau-distributed energy deposition of alpha particles traversing the air gap between the source and detector.

\begin{figure}[H]
  \centering
  \begin{subfigure}{0.49\textwidth}
    \centering
    \includegraphics[width=\textwidth]{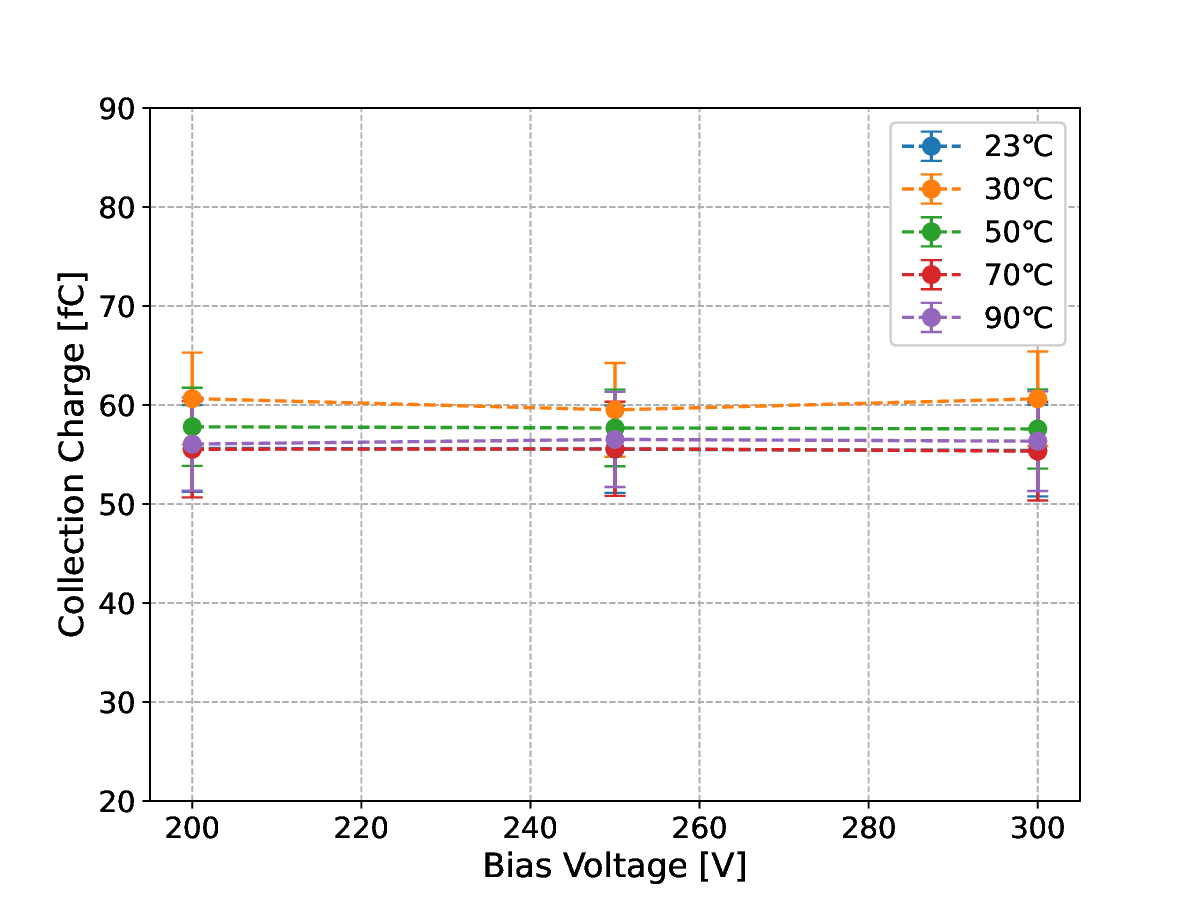} 
    \caption{}
  \end{subfigure}
  \hfill 
  \begin{subfigure}{0.49\textwidth}
    \centering
    \includegraphics[width=\textwidth]{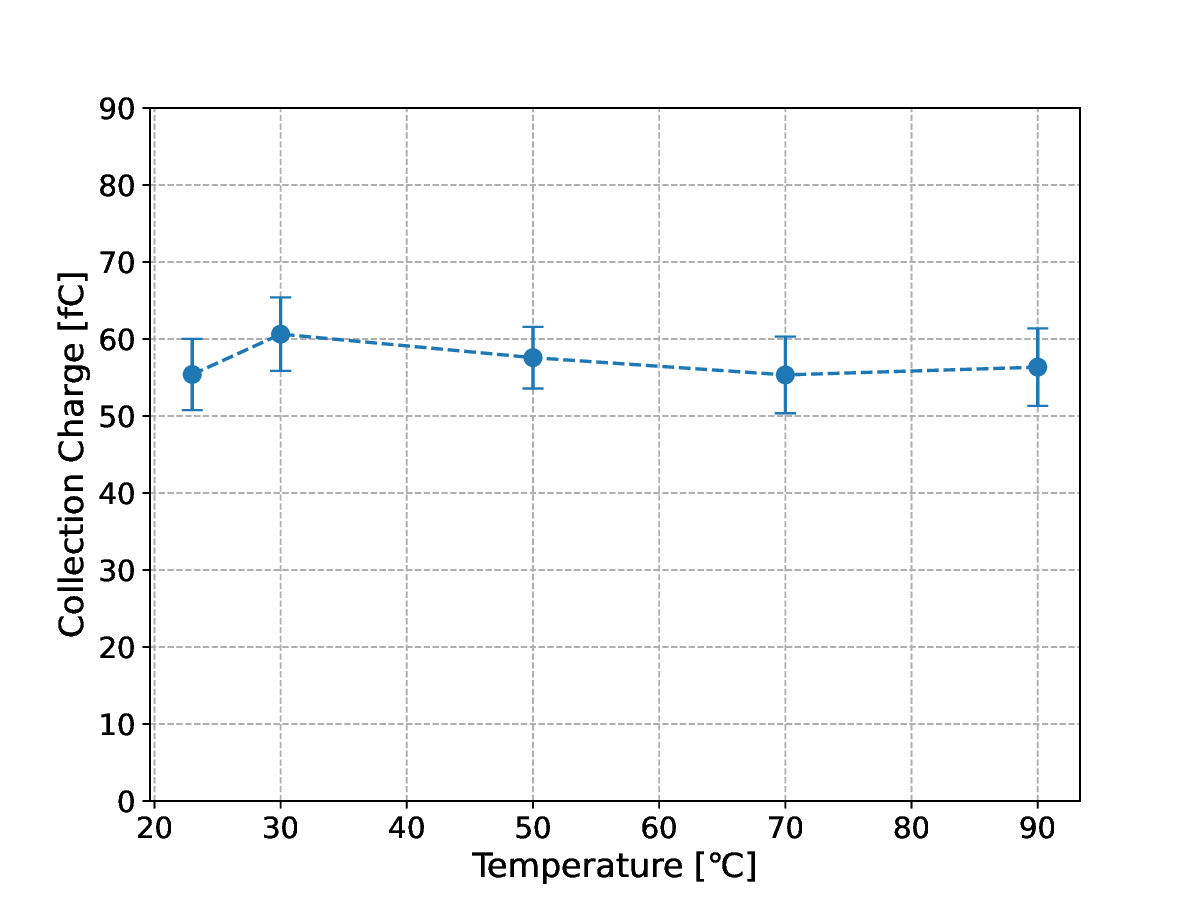} 
    \caption{}
  \end{subfigure}
  \caption{charge collection at (a) different voltage and (b) 300 V bias}
  \label{Charge_collection}
\end{figure}

Charge collection measurements were systematically performed across the temperature range from 23 °C to 90 °C at three bias voltages (200 V, 250 V, and 300 V). As shown in Figure \ref{Charge_collection}, the collected charge demonstrates good thermal stability, maintaining consistent values ($\pm ~ 10 \%$) throughout the entire temperature range. This temperature-independent performance confirms the exceptional stability of the SICAR PIN detector for high-temperature particle detection applications.

\begin{figure}[H]
\centering
\includegraphics[width=0.6\textwidth]{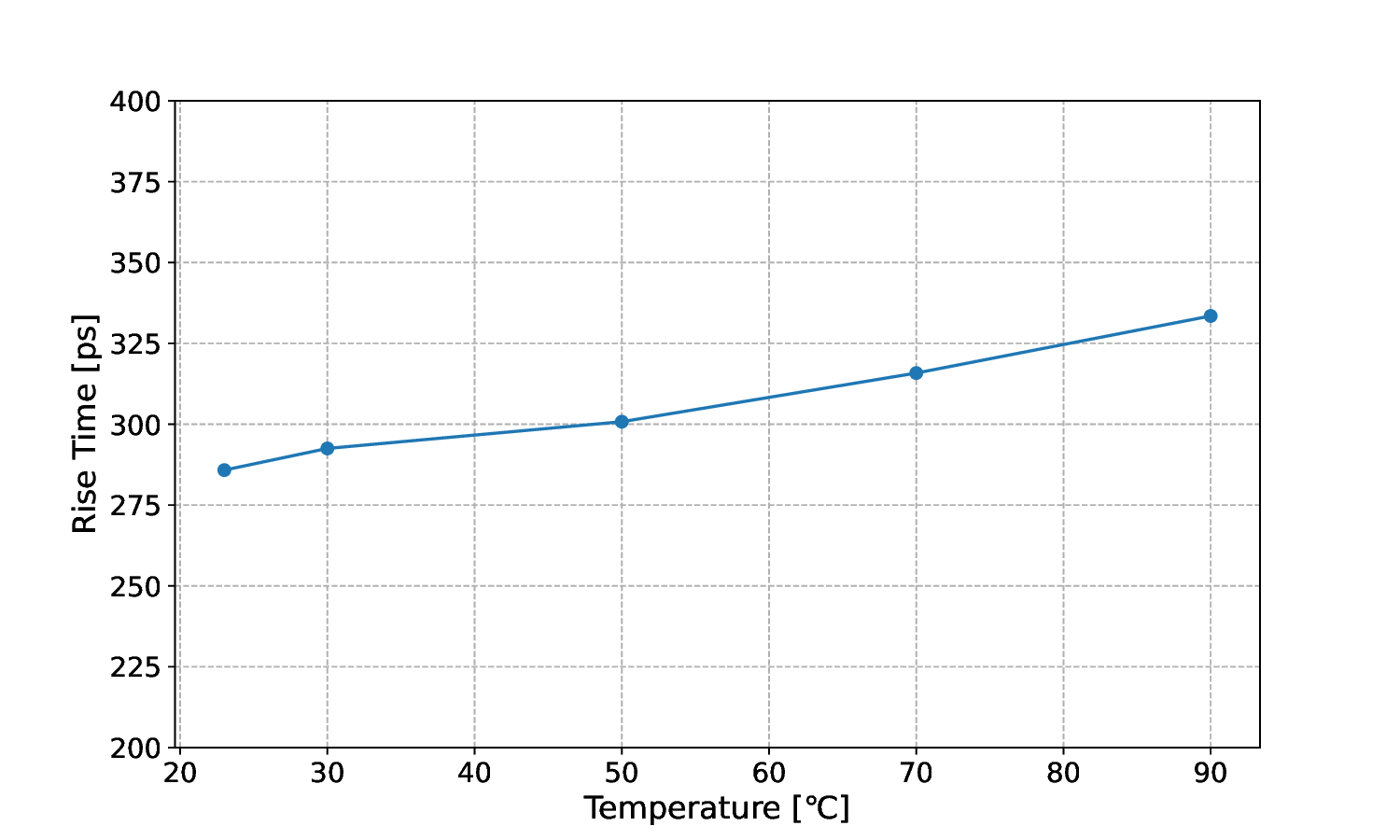}
\caption{rise time change with different temperature at 300 V}
\label{rise_time}
\end{figure}

In addition, the rise time change with temperature was also calculated. Taking the rise edge range from 10$\%$ to 90$\%$  of the signal’s amplitude as rise time value. As shown in Figure \ref{rise_time}, at 300 V bias, the rise time exhibits only slight temperature dependence, increasing gradually from 292.5 ps at 23 °C to 333.4 ps at 90 °C. This represents merely a 14$\%$ increase at elevated temperatures, demonstrating the detector's time resolution stability for potential high-temperature operation.

The excellent charge collection and fast rise time of SiC PIN devices at high temperatures establishes their strong potential for extreme-environment particle detection such as reactor monitoring. Moreover, they are expected to significantly reduce cooling costs for space exploration and further collider experiments.

\section{Conclusion}
A p-type silicon carbide PIN device with 30 $\rm \mu m$ active layer named SiCAR has been fabricated by IHEP. In this work, the annealing condition was optimized based on contact resistivity measurements using TLM method. The fabricated 4H-SiC PIN device with field plate and ring electrode structures maintains excellent operational stability across temperatures up to 90°C, exhibiting minimal leakage current (<10 nA at 300 V), stable depletion capacitance, consistent charge collection efficiency, and fast time response (333 ps rise time at 90°C). These results demonstrate that SiC PIN have great potential for demanding high temperature particle detection applications.

\section*{Acknowledgments}

This work is supported by the National Natural Science Foundation of China (Nos. 12205321, 12375184, 12305207, and 12405219), National Key Research and Development Program of China under Grant No. 2023YFA1605902 from the Ministry of Science and Technology, China Postdoctoral Science Foundation (2022M710085), Natural Science Foundation of Shandong Province Youth Fund (ZR2022QA098) under CERN RD50-2023-11 Collaboration framework and DRD3 SiC LGAD project.

\bibliographystyle{unsrt}

\end{document}